\documentclass[aps, prl, reprint, groupedaddress, superscriptaddress, amsmath, amssymb]{revtex4-1} 
\usepackage[colorlinks,allcolors=blue]{hyperref}
\usepackage{graphicx}
\usepackage{xcolor}
\usepackage{dcolumn}
\usepackage{bm}

\usepackage{braket}

\begin{document}
\title{Indirect but Efficient: Laser-Excited Electrons Can Drive Ultrafast Polarization Switching in Ferroelectric Materials}
\author{Chao Lian}
\author{Zulfikhar A. Ali}
\author{Hyuna Kwon}
\author{Bryan M. Wong}
\email{bryan.wong@ucr.edu}
\affiliation{Department of Chemical \& Environmental Engineering, Materials Science \& Engineering Program, and Department of Physics \& Astronomy, University of California-Riverside, Riverside, CA 92521, USA.}

\date{\today}

\begin{abstract}
\begin{minipage}[h]{0.55\linewidth}
To enhance the efficiency of next-generation ferroelectric (FE) electronic devices, new techniques for controlling ferroelectric polarization switching are required. While most prior studies have attempted to induce polarization switching via the excitation of phonons, these experimental techniques required intricate and expensive terahertz sources and have not been completely successful. Here, we propose a new mechanism for rapidly and efficiently switching the FE polarization via laser-tuning of the underlying dynamical potential energy surface. Using time-dependent density functional calculations, we observe an ultrafast switching of the FE polarization in BaTiO$_3$ within 200~femtoseconds. A laser pulse can induce a charge density redistribution that reduces the original FE charge order. This excitation results in both desirable and highly directional ionic forces that are always opposite to the original FE displacements. Our new mechanism enables the reversible switching of the FE polarization with optical pulses that can be produced from existing 800-nm experimental laser sources.\end{minipage} 
\hfill
\begin{minipage}[h]{0.2\linewidth}
\includegraphics[width=1.5\linewidth]{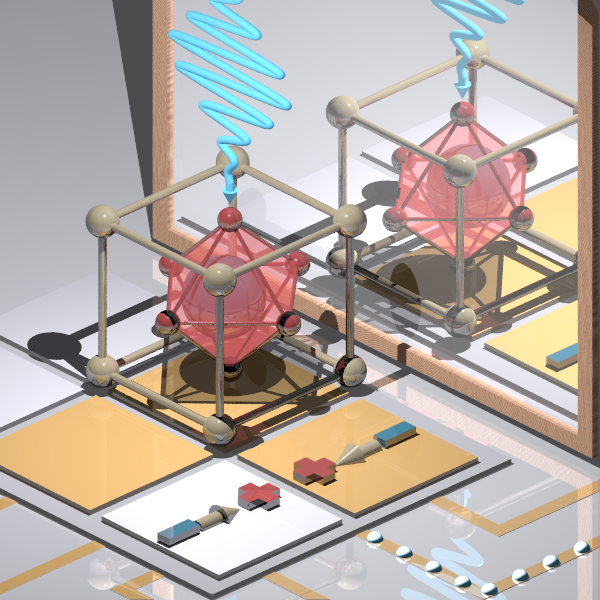}
\end{minipage}
\end{abstract}
\maketitle

\begin{figure}
\centering
\includegraphics[width=1\linewidth]{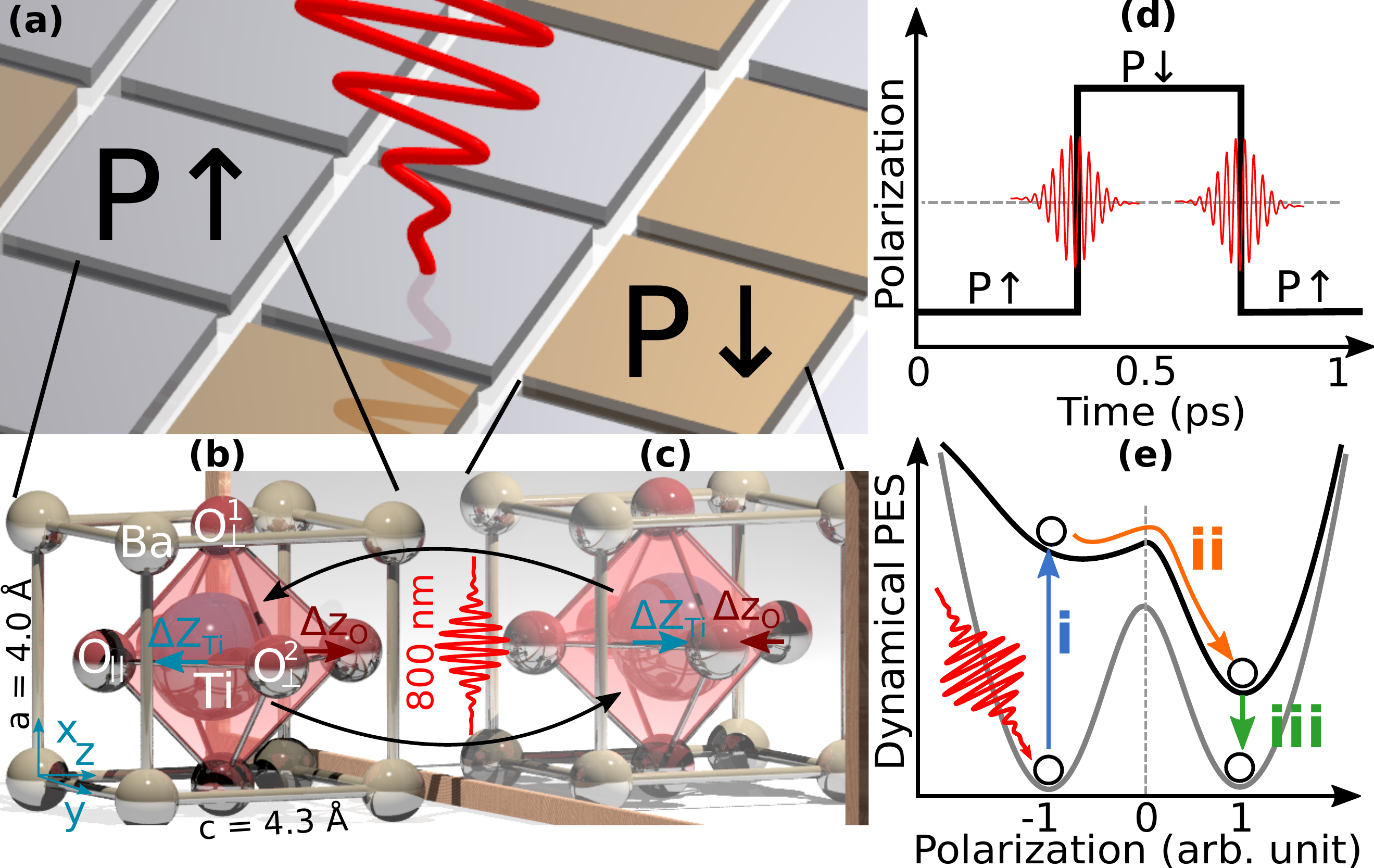}
\caption{(a) Schematic diagram of a FE array. The bright and dark blocks denote the (b) up-polarized (c) and down-polarized structures of BaTiO$_3$, respectively. (d) Diagram of ultrafast optical polarization switching as a function of time. The red lines denote two sequential identical laser pulses. (e) Diagram of laser-induced modification of the dynamical potential energy surface (PES). The gray (black) line represents the ground- and excited-state PES, respectively.}
\label{fig:Figure1}
\end{figure}

Ferroelectric (FE) materials are characterized by an intrinsic spontaneous electric polarization that can be further harnessed for next-generation electronic and energy-harvesting materials. For example, tuning the FE polarization can vary the tunneling resistance over several orders of magnitude~\cite{Dawber2005}, enabling technological advancements such as non-volatile memory in digital electronic devices,~\cite{Carlos1990} memristors,~\cite{Chanthbouala2012} and integrated neuromorphic networks~\cite{Jo2010, Kim2019}. In addition, by enabling a steady-state photocurrent, the use of FE polarization can substantially increase light-harvesting efficiency, particularly in hybrid organic-inorganic halide perovskite solar cells~\cite{Zhou2018,  Li2018d, Ghosh2015, Stroppa2015, Chen2014, Wei2014, Kim2015, Tong2017a, Ji2017, Jankowska2017b, Jankowska2017c, Li2017}. Finally, the variation in FE polarization on surfaces can also dramatically change the adsorption energetics in catalytic systems and could be further harnessed to enable other polarization-dependent surface mechanisms~\cite{Kakekhani2015, Bal2019}.

All of these applications are intrinsically associated with FE polarization and can, therefore, be further enhanced by tuning and controlling this intrinsic material property. Polarization switching is typically accomplished through a static electric field; however, the switching time is relatively sluggish (on the order of nanoseconds) due to the slow recrystallization time of most materials (typically hundreds of picoseconds)~\cite{Merz1954, Spierings2002}. To accelerate this switching process, significant research has focused on enabling ultrafast polarization switching via optical processes~\cite{Fahy1994, Li2004}. Terahertz (THz) sources provide ultrafast pulsed electric fields that are expected to reverse the polarization via the same mechanism as a static electric field. However, even with state-of-the-art THz sources, a single THz pulse does not possess enough strength to switch this polarization~\cite{Takahashi2006, Cavalleri2006, Qi2009a, Hauf2018, VonHoegen2018, Chen2016, Katayama2012}. To circumvent the use of THz sources, Subedi recently proposed that activating an infrared-active phonon mode can also induce polarization switching effects~\cite{Subedi2015}. Later, Mankowsky {et al.}~\cite{Mankowsky2017} conducted experiments on LiNbO$_3$ but only observed a temporary and partial polarization switching. While other researchers have reported related phenomena such as ultrafast domain wall movements,~\cite{Daranciang2012, Lejman2014, Ahn2017, Akamatsu2018, Shinde2018} photoinduced depolarization,\cite{KorffSchmising2007, Rana2009, Kuo2017a, Rubano2018} and coherent ionic movement~\cite{Forst2013, Porer2018, Qi2018}, ultrafast polarization switching in these FE materials have not been completely successful. Ideally, one would like a more efficient mechanism, such as that shown in Fig.~\ref{fig:Figure1}(a) and (d), in which a readily-available laser source could be utilized to switch the polarization with one pulse and reversibly switched back with a second identical pulse. 

In this Letter, we propose a new mechanism that enables this efficient switching of the polarization in FE materials. By accounting for the detailed interactions between the electrons, nuclei, and electromagnetic field, our real-time time-dependent density functional theory (RT-TDDFT) simulations show that the ultrafast FE polarization switching in BaTiO$_3$ occurs within 200~femtoseconds (fs), after the material is excited by an 800-nm laser pulse. We find that this polarization switching commences when the laser pulse pumps electrons from 2p orbitals in oxygen to the 3d orbitals in titanium. This dynamic excitation process yields desirable and highly-directional forces on the Ti and O atoms that are (1) always opposite to the original FE displacements and (2) can be harnessed to consistently switch the polarization of the material. By laser-tuning the dynamical potential energy surface, we show that this new mechanism can switch the FE polarization in both directions using identical pulses from experimentally-available laser sources.

We use our in-house time-dependent \textit{ab initio} package, (\textsf{TDAP})~\cite{Meng2008a, Lian2018MultiK, Lian2018AdvTheo}, for our RT-TDDFT calculations~\cite{Runge1984,Bertsch2000,Wang2015a}, where the wavefunctions and charge densities are obtained from the \textsf{Quantum Espresso}~\cite{Giannozzi2009, Giannozzi2017} software package. We used the projector augmented-waves method (PAW)~\cite{Blochl1994} and the Perdew-Burke-Ernzerhof (PBE) exchange-correlation (XC) functional~\cite{Perdew1996} in both our DFT and RT-TDDFT calculations. Pseudopotentials were generated using the \textsf{pslibrary}~\cite{DalCorso2014} software package. The plane-wave energy cutoff was set to 55~Ry, and the Brillouin zone was sampled using a Monkhorst-Pack scheme with an $8\times8\times8$ $k$-point mesh for the unit cell and a $2\times2\times2$ mesh for a $3\times3\times3$ supercell. To reproduce the experimental band gap, a scissor correction of 1.65~eV was added to both the ground state and time-dependent calculations. The electronic timestep, $\delta t$, was set to $1.94\times10^{-4}$~fs, and the ionic timestep, $\Delta t$, was $0.194$~fs. The Gaussian-type laser pulse utilized in our study is given by
    $
    \label{eq:GaussianWave}
    \mathbf{E}(t)=\mathbf{E}_0\cos\left(\omega t \right)\exp\left[-\frac{(t-t_0)^2}{2\sigma^2}\right],
    $
where $|\mathbf{E}_0|$ is the electric field amplitude, $\omega=1.55$~eV is the laser frequency, and $t_0=50$~fs is the temporal location of the electric field peak. The laser pulse is linearly polarized along the $x$ direction, perpendicular to the ferroelectric polarization. The crystal orbital Hamilton population (COHP) analysis was calculated with the \textsf{Lobster}~\cite{Dronskowski1993, Deringer2011, Maintz2016} software package. 
\begin{table}
\caption{\label{tab:struct} Symmetric positions $\alpha^{\mathrm{sym}}_I$ ($\alpha=x, y, z$), FE displacements $\Delta z_I$ and Born effective charge of the $I$-th atom. We categorize these three O atoms into two types: one O$_{\parallel}$ atom that is parallel to the polarization and two O$_{\perp}$ atoms that are perpendicular to the polarization.}
\begin{ruledtabular}
\begin{tabular}{llllrrr}
	   Atom $I$     & $x^{\mathrm{sym}}_I$ & $y^{\mathrm{sym}}_I$ & $z^{\mathrm{sym}}_I$ & $\Delta z$ (\AA) & $\mu^*$ ($|e|$) &  \\\hline
	      Ba        &         0.0         &         0.0         &         0.0         &      0.00       &      2.73      &  \\
	      Ti        &        0.5a         &        0.5a         &         0.5         &      0.08       &      7.17      &  \\
	 O$^1_{\perp}$  &         0.0         &        0.5a         &        0.5c         &      -0.12      &     -2.02      &  \\
	 O$^2_{\perp}$  &        0.5a         &         0.0         &        0.5c         &      -0.12      &     -2.02      &  \\
	O$_{\parallel}$ &        0.5a         &        0.5a         &         0.0         &      0.21       &     -5.74      &  \\
\end{tabular}
\end{ruledtabular}
\end{table}

We first optimized the structure of BaTiO$_3$, which is shown in Fig.~\ref{fig:Figure1}(b) and Table~\ref{tab:struct}. The lattice is tetragonal with $a=b=4.00$~\AA\ and $c/a = 1.075$. The ferroelectric properties of BaTiO$_3$ originate from the slight distortion of the Ti-O octahedron: the Ti atom deviates from the body-centered position along the $z$ direction and the O atoms deviate from the face-centered positions along the $-z$ direction. These FE displacements are characterized as the displacements from the symmetric positions $\Delta \alpha_I = \alpha^{\mathrm{FE}}_I-\alpha^{\mathrm{sym}}_I$, where $\alpha^{\mathrm{FE}}_I$ and $\alpha^{\mathrm{sym}}_I$ are the positions of atom $I$ ($I = $ Ti, Ba, O$_\perp$, and O$_\parallel$) along the $\alpha = x$, $y$, and $z$ direction in the FE and symmetric phase, respectively. We calculate the FE polarization accordingly as $P_0 = \frac{1}{V}\sum_I \mu^*_I \Delta z_I = 2.01~\mathrm{e/\AA^2} = 0.50~\mathrm{C/m^2}$, where $V=a\times a\times c=68.8$~\AA$^3$ is the volume, $\Delta z_I$ is the ionic FE displacement, and $\mu_I^*$ is the Born effective charge, as shown in Fig.~\ref{fig:Figure1}(b) and Table~\ref{tab:struct}. Consistent with previous studies, the semilocal exchange-correlation functionals slightly overestimate the $c/a$ ratio and the static polarization, as discussed in \cite{Bilc2008, Sun2016}.

\begin{figure}
\centering
\includegraphics[width=1.0\linewidth]{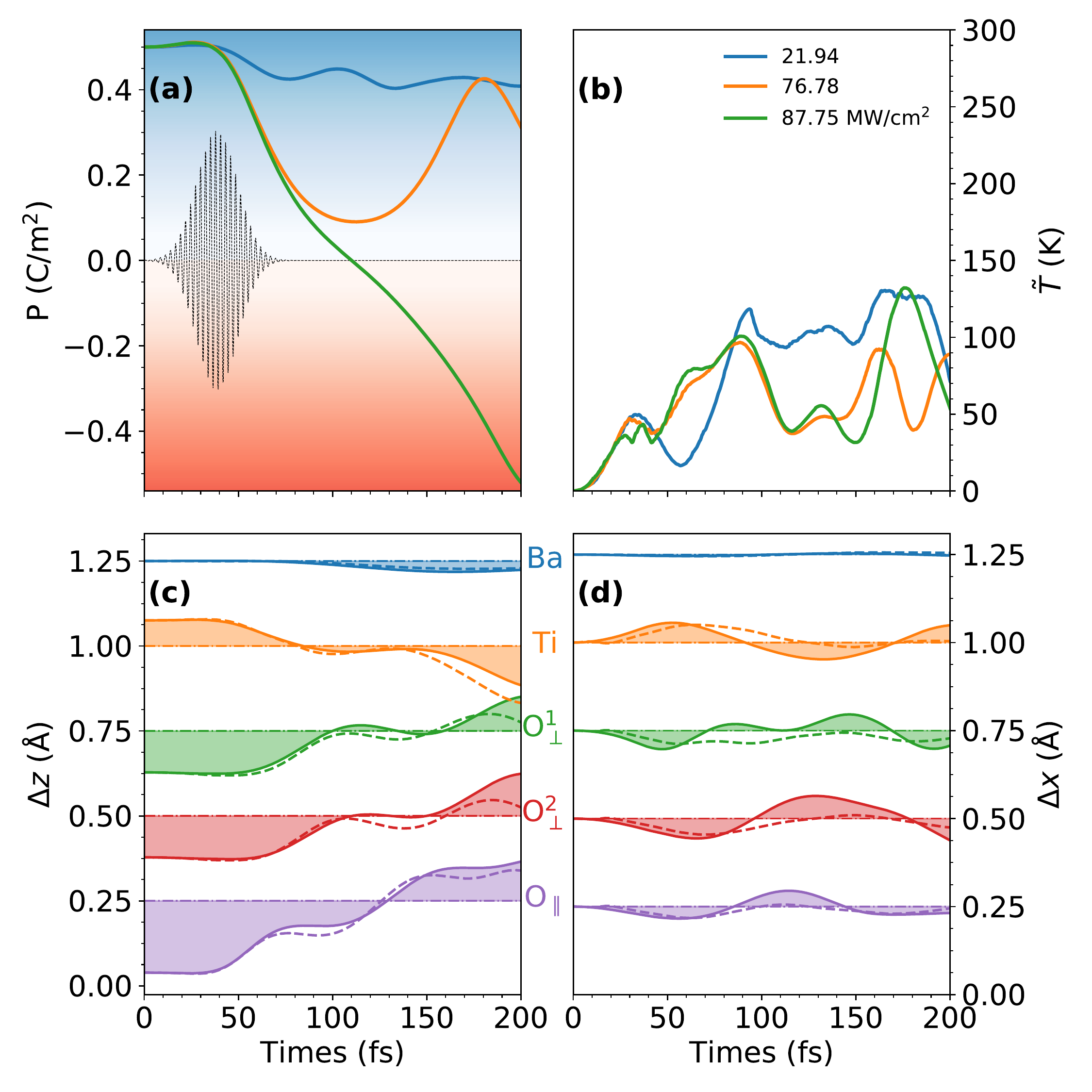}
\caption{(a) FE polarization $P$ and (b) Effective temperature $\tilde{T}$ as a function of time. FE displacements as a function of time for (c) $\alpha=z$ (d) $\alpha=x$. Solid lines denote the FE displacements calculated in the $1\times1\times1$ cell, and dashed lines denote the average FE displacements in the $3\times3\times3$ cell.
}
\label{fig:struct}
\end{figure}
We apply laser pulses having a wavelength of $\lambda = 800$~nm, a polarization perpendicular to the FE polarization direction, and a duration lasting from $t_0=0$~fs to $t_f=100$~fs. We characterize and monitor the laser-induced dynamics by calculating the dynamical polarization
\begin{equation}
P(t) = \frac{1}{V}\sum_I \mu^*_I \Delta z_I(t).
\end{equation}
Figure~\ref{fig:struct}(a) shows the results under various laser fluences $F=$ 21.94, 76.78, and 87.75~MW/cm$^2$. We identify three types of lattice dynamics: \textbf{(1)} at the lowest fluence of $F=21.94$~MW/cm$^2$, $P(t)$ slightly decreases and oscillates around $P_0$. \textbf{(2)} when $F$ increases to a critical fluence of 76.78~MW/cm$^2$, $P(t)$ substantially decreases to around 0 within 100~fs, while $P(t)$ recovers to $P_0$ in the following 100~fs. \textbf{(3)} When $F$ further increases to 87.75~MW/cm$^2$, we observe a switching of the polarization. $P(t)$ continuously decreases past 0 until it reaches its maximum polarization in the opposite direction.

In contrast to other experimental strategies discussed previously, this laser-induced polarization change is a non-thermal process. To clearly demonstrate this, we plot the effective temperature $\tilde{T}(t) = E_{\mathrm{kin}}(t)/k_B$ as a function of time, where $E_{\mathrm{kin}}$ is the kinetic energy of the ions and $k_B$ is the Boltzmann constant. As shown in Fig.~\ref{fig:struct}(b), $\tilde{T}(t)$ is much lower than room temperature throughout our entire dynamics simulations regardless of the laser fluence. In addition, $\tilde{T}(t)$ is different from the thermodynamic equilibrium temperature, which is produced by the random ionic movements. Instead, in the laser-induced polarization switching process, the ionic movements are highly directional. As shown in Fig.~\ref{fig:struct}(c), the laser radiation triggers a set of FE displacements, $\Delta z_I(t)$, that are nearly anti-parallel with the original values, $\Delta z_I(t_0)$. $\Delta z_I(t)$ of the oxygen atoms $O_\perp$ and $O_\parallel$ changes sign from negative to positive, while the titanium atom moves in the opposite direction from positive to negative. Considering that the oxygen and titanium atoms have effective Born charges with opposite signs (Table~\ref{tab:struct}), the movements of all the atoms contribute consistently to the polarization switching. Moreover, along the directions perpendicular to the polarization, the displacement, $\Delta x_I(t)$, just slightly oscillates around the equilibrium position, as shown in Fig.~\ref{fig:struct}(d). This leads to an abnormal fluence-dependence of the effective temperature $\tilde{T}(t)$, which slightly decreases as the fluence increases. Although the laser pulse induces larger ionic forces at higher fluence, the system needs to overcome the small energy barrier to reverse the polarization, as shown in Fig.~\ref{fig:Figure1}(e), which consumes the kinetic energy and decreases the effective temperature. In comparison, with a lower influence, the ions can not overcome the energy barrier, resulting in a higher effective temperature for the oscillation. Thus, the highly directional FE movements predominantly contribute to the relatively low effective temperature $\tilde{T}(t)$, indicating a unique laser-induced non-thermal mechanism.

Coherence is another distinguishable feature of our polarization-switching mechanism in which the laser pulse drives the movement of the ions in a synchronized fashion. To verify that this coherence is not an artifact of size effects due to the use of a single unit cell, we carried out the TDDFT calculations using a $3\times3\times3$ supercell for comparison. We analyzed the average laser-induced displacements as a function of time, $\left<\Delta{z_I}(t)\right>= \sum_{s=1}^{27} \Delta z^s_I(t)$, where $s$ denotes the index of the unit cell. As shown in Fig.~\ref{fig:struct}(c) and (d), $\left<\Delta{z_I}(t)\right>$ in the $3\times3\times3$ supercell case is almost identical to $\Delta z_I(t)$ in the $1\times1\times1$ unit cell case, which indicates that the unified switching of the FE displacements occurs over all the laser-irradiated area. Consequently, no nucleation and growth of oppositely polarized domains are needed in this polarization switching mechanism, which accelerates this switching process to finish within 200~fs.

\begin{figure}
\centering
\includegraphics[width=1\linewidth]{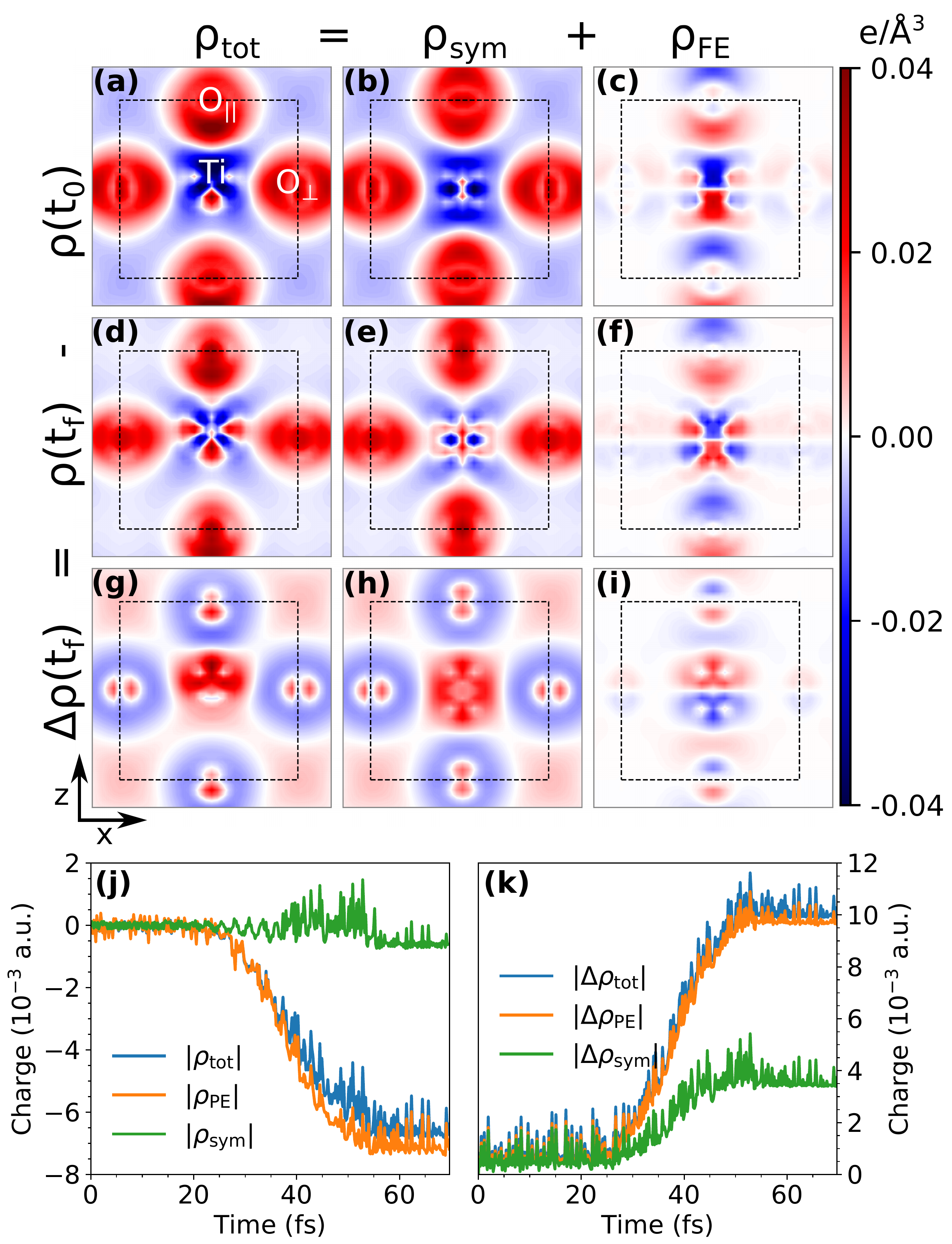}
\caption{Two-dimensional contour plots of the charge densities at the plane $y=a/2$ for (a) $\rho_{\mathrm{tot}}(x,z,t_0)$, (b) $\rho_{\mathrm{sym}}(x,z,t_0)$, (c) $\rho_{\mathrm{FE}}(x,z,t_0)$, (d) $\rho_{\mathrm{tot}}(x,z,t_f)$, (e) $\rho_{\mathrm{sym}}(x,z,t_f)$, (f) $\rho_{\mathrm{FE}}(x,z,t_f)$, (g) $\Delta\rho_{\mathrm{tot}}(x,z,t_f)$, (h) $\Delta\rho_{\mathrm{sym}}(x,z,t_f)$, and (i) $\Delta\rho_{\mathrm{FE}}(x,z,t_f)$, where $t_0$ and $t_f$ are the start and end times of the laser pulse, respectively. Panels (j) and (k) depict the bonding and anti-bonding charge, respectively, as a function of time. For ease of comparison, the initial values of $\rho_{\mathrm{tot}}$, $\rho_{\mathrm{sym}}$, and $\rho_{\mathrm{FE}}$ in (j) are downshifted by $20.7\times10^{-3}$, $20.0\times10^{-3}$, and $5.70\times10^{-3}$ a.u., respectively.}
\label{fig:2Dimage+orderdyn}
\end{figure}
We illustrate the atomistic mechanisms of the polarization switching by analyzing the carrier dynamics: i.e., we monitor the differential charge density $\rho_{\mathrm{tot}}(x,y,z,t) = \rho_{\mathrm{chg}}(x,y,z,t) - \rho_{\mathrm{atom}}(x,y,z, t)$ as a function of time, where $\rho_{\mathrm{chg}}$ is the charge density, and $\rho_{\mathrm{atom}}$ is the superposition of the atomic charge densities. Thus, $\rho_{\mathrm{tot}}(x,y,z,t)$ captures the spatial distribution of the bonding (+) and antibonding (--) densities. The FE polarization can be treated as the asymmetric disturbance of the symmetric phase, which is characterized by the centrosymmetric charge order. Thus, we divide the total charge $\rho_{\mathrm{tot}}(x,y,z,t)$ into two parts, $\rho_{\mathrm{sym}}(x,y,z)$ and $\rho_{\mathrm{FE}}(x,y,z)$, where $\rho_{\mathrm{sym}}(x,y,z) = [\rho_{\mathrm{tot}}(x,y,c-z) + \rho_{\mathrm{tot}}(x,y,z)]/2$, and $\rho_{\mathrm{FE}}(x,y,z) = \rho_{\mathrm{tot}}(x,y,z) - \rho_{\mathrm{sym}}(x,y,z)$. Obviously, $\rho_{\mathrm{sym}}(x,y,\frac{1}{2}c-z) = \rho_{\mathrm{sym}}(x,y,\frac{1}{2}c+z)$ is the centrosymmetric part of the total charge that characterizes the original symmetric order, whereas $\rho_{\mathrm{FE}}(x,y,z)$ is the asymmetric part that characterizes the FE order.

Figure~\ref{fig:2Dimage+orderdyn}(a)-(i) depicts a two-dimensional contour of the charge density, $\rho_i(x,z,t)$ ($i=\mathrm{tot}$, $\mathrm{sym}$, and $\mathrm{FE}$), at the Ti-O surface $y=a/2$. Comparing Fig.~\ref{fig:2Dimage+orderdyn}(a)-(c), we find that $\rho_{\mathrm{sym}}(x,z,t_0)$ is the dominant component of $\rho_{\mathrm{tot}}(x,z,t_0)$ even when FE effects are included, whereas $\rho_{\mathrm{FE}}(x,z,t_0)$ is localized around the Ti positions. As such, these figures show that an electron transfers from a Ti 3d orbital to an O 2p orbital, which triggers a pseudo-Jahn-Teller effect (PJTE)~\cite{Polinger2015} and stabilizes the FE phase. By comparing the laser-induced charge density difference, $\Delta\rho_i(x,z,t) = \rho_i(x,z,t) - \rho_i(x,z,t_0)$ ($i=\mathrm{tot}$, $\mathrm{sym}$, $\mathrm{FE}$), with the ground state charge density $\rho_i(x,z,t_0)$, we find that: (1) the laser induces an electron transfer from a bonding to an anti-bonding area, where $\Delta\rho_\mathrm{tot}(x,z,t_f)$ [Fig.~\ref{fig:2Dimage+orderdyn}(g)] is opposite to that of the ground-state bonding charge $\rho_\mathrm{tot}(x,z,t_0)$ [Fig.~\ref{fig:2Dimage+orderdyn}(a)]; (2) the majority of the induced charge has a centrosymmetric periodicity, i.e. $\Delta\rho_\mathrm{sym}(x,z,t_f)$ [Fig.~\ref{fig:2Dimage+orderdyn}(h)] dominates the induced total charge $\Delta\rho_\mathrm{tot}(x,z,t_f)$ [Fig.~\ref{fig:2Dimage+orderdyn}(g)]; (3) The most important feature is that the induced FE charge density $\Delta\rho_{\mathrm{FE}}(x,z,t_f)$ [Fig.~\ref{fig:2Dimage+orderdyn}(i)] is opposite to that of the original $\rho_{\mathrm{FE}}(x,z,t_0)$ [Fig.~\ref{fig:2Dimage+orderdyn}(c)], indicating a decrease in the FE order. We quantitatively evaluate the laser-induced change in the charge density by analyzing the integrated charges $Q_{i}(t) = \int |\rho_i(\mathbf{r},t)| d\mathbf{r}$ and $C_{i}(t) = \int |\rho_i(\mathbf{r}, t) - \rho_i(\mathbf{r}, t_0)| d\mathbf{r} = \int |\Delta\rho_i(\mathbf{r},t)| d\mathbf{r}$, where $i=\mathrm{tot}$, $\mathrm{sym}$, and $\mathrm{FE}$. The former characterizes the change in the bonding strength and the latter denotes the weakening of bonds, respectively. As shown in Fig.~\ref{fig:2Dimage+orderdyn}(j), the percentage decrease in the bond strength, $[Q_{i}(t_f)-Q_{i}(t_0)]/Q_{i}(t_0)$, is 32.45\%, 35.92\%, and 10.77\% for $i=\mathrm{tot}$, $\mathrm{sym}$, and $\mathrm{FE}$, respectively. Accordingly, $C_{i}(t)$ increases, with a ratio of $C_{\mathrm{tot}}(t_f):C_{\mathrm{sym}}(t_f):C_{\mathrm{FE}}(t_f) = 1:0.97:0.27$, which is nearly the same as the ratio of the initial bonding charges $Q_{\mathrm{tot}}(t_0):Q_{\mathrm{sym}}(t_0):Q_{\mathrm{FE}}(t_0) = 1:0.97:0.34$. Thus, the laser-induced bonding-antibonding transfer is nearly homogeneous, lowering both the centrosymmetric and FE order proportionally. Since the decrease in $Q_{\mathrm{sym}}$ affects all chemical bonds homogeneously, the overall effect of the decrease in $Q_{\mathrm{tot}}$ is to lower the stability of the FE polarization. Thus, the laser-induced charge-density change, $\Delta\rho_\mathrm{tot}(x,z,t)$ is always opposite to the original FE charge order and, therefore, weakens the original ionic bonding and PJTE, creating the unsymmetrical dynamical PES shown in Fig.~\ref{fig:Figure1}(e).

\begin{figure*}
	\centering
	\includegraphics[width=1.0\linewidth]{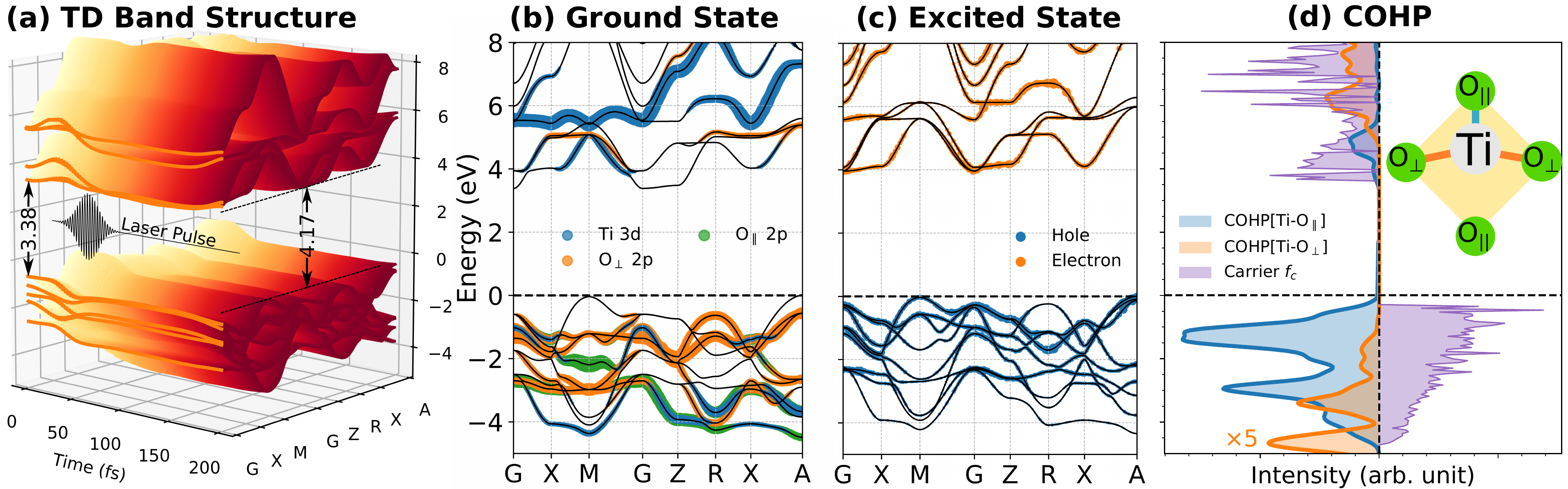}
	\caption{(a) Time-dependent band structures, and snapshot of band structures at (b) $t=0$~fs and (c) $t=150$~fs. The circles in (b) denote the momentum-resolved projected density of states (PDOS) of the Ti 3d and O 2p orbitals. The circles in (c) denote the momentum-resolved distribution of photocarriers. Panel (d) depicts the crystal orbital Hamilton population (COHP) of Ti-O bonds and the photocarrier distribution $f_c$ as a function of energy. The COHP of Ti-O$_{\perp}$ is multiplied by 5 for comparison. The negative (positive) region of the COHP denotes bonding (antibonding) states. The negative (positive) values of the photocarrier distribution denote the photoinduced electron (hole) density.}
	\label{fig:Figure4}
\end{figure*}

We now analyze the carrier dynamics in momentum space for the inter-orbital transition. As shown in Fig.~\ref{fig:Figure4}(a), the time-dependent band structures from our RT-TDDFT calculations show that the band gap increases from 3.38 to 4.17~eV while the band dispersions only slightly change. Since the initial and final states are degenerate at the ground state, the gap increase is attributed to the photocarrier generation. Comparing Fig.~\ref{fig:Figure4}(b) and (c), the multi-photon excitation process produces carriers that are distributed from -4 to 6 eV, which indicates that the laser-induced electrons are mostly transfered from O$_{2p}$ orbitals to Ti$_{3d}$ orbitals. This transition is opposite to the electron flow induced by the PJTE, which introduces a highly-selective bond weakening for Ti-O$_{\parallel}$ over the Ti-O$_{\perp}$ bond. We can quantitatively evaluate the weakening of Ti-O bonds as 
\begin{equation}
\Delta\eta[\mathrm{bond}] = \int_{-\infty}^{\infty} \mathrm{COHP}[\mathrm{bond}](\epsilon)f_c(\epsilon) d\epsilon,
\end{equation}
where $\epsilon$ is the energy, $\mathrm{bond}$ represents either the $\mathrm{Ti-O}_{\perp}$ or $\mathrm{Ti-O}_{\parallel}$ bond, $\mathrm{COHP}(\epsilon)$ denotes the crystal orbital Hamilton population (COHP), and $f_c(\epsilon)$ is the photocarrier distribution, as shown in Fig.~\ref{fig:Figure4}(d). From this analysis, we find that $\Delta\eta[\mathrm{Ti-O}_{\parallel}] = 1.57$ is much larger than $\Delta\eta[\mathrm{Ti-O}_{\perp}] = 0.01$, which indicates that the weakening of $\mathrm{Ti-O}_{\parallel}$ is much more significant than the weakening of the $\mathrm{Ti-O}_{\perp}$ bond. Thus, the anisotropic Ti-O bonds in BaTiO$_3$ result in a highly directional laser-induced polarization instability: the bonds parallel to the polarization are largely weakened while the bonds perpendicular to the polarization direction are barely affected.

Based on our photocarrier dynamic analysis of BaTiO$_3$, we have uncovered an electronically-driven polarization switching mechanism that is different from the conventional ionic-driven mechanism used in existing THz experiments. As shown in Fig.~\ref{fig:Figure1}(e), our mechanism is based on a laser-tuned dynamical potential energy surface (PES), which can be generated with the time-dependent energy $E(t)$ and the time-dependent polarization $P(t)$ obtained in the TDDFT calculation as $E[P(t)]$. Since the dynamical PES is time-dependent, we only show the diagram to simplify the illustration. The laser radiation transforms the ground state PES into a dynamical PES by first exciting the material and automatically driving the ions to the opposite polarization. This PES-based mechanism encompasses three steps: \textbf{(1)} the laser pulse raises the PES of BaTiO$_3$ by pumping electrons from the 2p orbitals of oxygen to the 3d orbitals of titanium, inducing an anti-FE charge order [Fig.~\ref{fig:2Dimage+orderdyn}(i)]; \textbf{(2)} this dynamic excitation process yields desirable and highly-directional forces on the Ti and O atoms that are always opposite to the original FE displacements, transforming the lattice into a structure with an opposite FE polarization along the non-equilibrium TD-PES within 200~fs [Fig.~\ref{fig:struct}(a)]; \textbf{(3)} the photo-excited system relaxes to the ground state in conjunction with the recombination of the photo-carriers (which occurs beyond our simulation time). We speculate that this dynamical-PES-based polarization switching is ubiquitous in all the FE materials and, therefore, the same mechanism can be used to manipulate the polarization in other FE perovskite materials such as PbTiO$_3$ and LiNbO$_3$, multiferroic materials, and even recently-discovered two-dimensional SnTe-based FE materials.~\cite{Chang2016} Most importantly, this new mechanism enables the switching of the FE polarization in both directions, with identical pulses that can be produced from experimentally-available 800-nm laser sources. Taken together, these findings provide a new mechanistic understanding of electronically-driven FE dynamics via laser-tuning of the dynamical potential energy surface, which can accelerate the design of more efficient, ultrafast FE devices.

C. L. and H. K. acknowledge support from the UC Riverside Collaborative Seed Grant. Z. A. A. and B. M. W. acknowledge financial support from the Office of Naval Research (Grant N00014-18-1-2740).

%

\end{document}